\documentclass[12pt]{article}
\usepackage{epsfig}
\begin{document}
%%%%%%%%%%%%%%%%%%%%%%%%%%%%%%%%%%%%%%%%%%%%%%%%%%%%%%%%%%%%%%%%%%%%%%%%%
\title{The role of hidden-charm pentaquark resonance $P^+_c(4450)$\\
                 in $J/\psi$ photoproduction on nuclei near threshold}
\author{E. Ya. Paryev$^{1,2}$ and Yu. T. Kiselev$^2$\\
{\it $^1$Institute for Nuclear Research, Russian Academy of Sciences,}\\
{\it Moscow 117312, Russia}\\
{\it $^2$Institute for Theoretical and Experimental Physics,}\\
{\it Moscow 117218, Russia}}
%==============================================================
%%==============================================================

\renewcommand{\today}{}
\maketitle

\begin{abstract}
   We study the $J/\psi$ photoproduction from nuclei at near-threshold incident photon energies of 5--11 GeV within
   the nuclear spectral function approach by considering incoherent direct (${\gamma}N \to {J/\psi}N$) and two-step
   (${\gamma}p \to P^+_c(4450)$, $P^+_c(4450) \to {J/\psi}p$) photon--nucleon charmonium creation processes.
   We calculate the absolute and relative excitation functions within the different scenarios for in-medium
   modification of the directly photoproduced $J/\psi$ mesons. We show that the overall subthreshold $J/\psi$
   production in ${\gamma}A$ reactions reveals some sensitivity to adopted in-medium modification scenarios
   for $J/\psi$ mesons only if branching ratio $Br[P^+_c(4450) \to {J/\psi}p]$ $\sim$ 1\% and less.
   Our studies also demonstrate that the presence of the $P^+_c(4450)$ resonance
   in $J/\psi$ photoproduction produces above threshold additional enhancements in the
   behavior of the total $J/\psi$ creation cross section on nuclei, which could be also studied in the future
   JLab experiments at the upgraded up to 12 GeV CEBAF facility to provide further evidence for its existence.
\end{abstract}

\newpage

\section*{1. Introduction}

In a recent publication [1] the $J/\psi$ meson photoproduction on nuclei was investigated in the framework
of the spectral function approach based on considering the direct non-resonant (or the so-called elastic)
processes ${\gamma}N \to {J/\psi}N$ of $J/\psi$ production at near-threshold
\footnote{$^)$The $J/\psi$ threshold photoproduction energy on a free nucleon at rest is 8.2 GeV.}$^)$
incident photon energies of 5--11 GeV within the different scenarios for its in-medium mass shift at saturation
density $\rho_0$. As a measure for this shift the initial photon energy dependence of $J/\psi$ production
cross sections on nuclei (excitation functions) has been employed in this work. It has been shown that these
functions reveal definite sensitivity to the possible charmonium in-medium mass shifts in the range of -50 MeV
and more (in absolute magnitude) at far subthreshold beam energies $\sim$5-7 GeV, while smaller mass shifts
will probably be experimentally inaccessible. The above conclusion was drawn by neglecting in part for
${\gamma}A$ reactions the $J/\psi$ creation in the two-step elementary processes
${\gamma}p \to P_c^+$, $P_c^+ \to {J/\psi}p$
with an intermediate hidden-charm pentaquark resonances, $P^+_c(4380)$ and $P^+_c(4450)$, recently discovered
by the LHCb Collaboration in the invariant mass spectrum of ${J/\psi}p$ in the decay
$\Lambda^0_b \to K^-({J/\psi}p)$ [2]. These processes may also contribute to the $J/\psi$ production in
near-threshold ${\gamma}A$ interactions, since the laboratory photon energies of 9.75 and 10.08 GeV are
needed to excite the $P^+_c(4380)$ and $P^+_c(4450)$ resonances with masses of 4.38 and 4.45 GeV [2],
respectively, on a free target proton at rest. Their detailed studying is planned to be performed at
JLab 12 GeV in the near future, using the SoLID detector in Hall A [3], the CLAS12
detector in Hall B [4] as well as the HMS and SHMS detectors in Hall C [5], with detecting in part the $J/\psi$
meson via its $J/\psi \to e^+e^-$ decay to provide both further evidence of the LHCb pentaquark states existence
and to investigate the internal structure of such states. This structure is differently interpreted in the
literature, namely: in terms of quark degrees of freedom [6--10], as a hadronic molecule consisting of a charmed
baryon and charmed meson [11--16], as a bound state of the charmonium $\psi(2S)$ and the proton [17],
$\chi_{c1}$ and the proton [18]
\footnote{$^)$It should be noticed that a recent work [19] disfavors the previous interpretation
[20] of the LHCb peak $P^+_c(4450)$, having the quantum numbers preferred by the experimentalists,
in terms of a kinematic triangle singularity in the $\Lambda^0_b \to K^-({J/\psi}p)$ decay.}$^)$
.

  In addition to the above experimental plans, it is interesting to clarify the role of the presumed
pentaquark resonances $P^+_c(4380)$ and $P^+_c(4450)$ in near-threshold $J/\psi$ photoproduction off nuclei
to get a deeper insight into the possible modification of the $J/\psi$ meson mass in nuclear medium and into
the feasibility of its determination experimentally.
The primary goal of the present work is to extend the model [1] to $J/\psi$--producing two-step resonant
process ${\gamma}p \to P_c^+(4450)$, $P_c^+(4450) \to {J/\psi}p$ with the highest mass narrow exotic
resonance $P^+_c(4450)$ in the intermediate state
\footnote{$^)$We will not consider in the present study the resonant elementary two-step process
${\gamma}p \to P_c^+(4380)$, $P_c^+(4380) \to {J/\psi}p$, since its total cross section is expected to be
less than that of the channel ${\gamma}p \to P_c^+(4450)$, $P_c^+(4450) \to {J/\psi}p$ by a factor of
about ten [21--23].}$^)$
.
In what follows, we briefly recall the main assumptions of the model [1] and describe the respective extension.
Also we present both the predictions [1] and those obtained within this extended model for the $J/\psi$
excitation functions in ${\gamma}C$ and ${\gamma}Pb$ collisions near threshold.
These predictions can be used as an important tool for possible extraction of valuable information on the
charmonium in-medium mass shift from the data which could be taken in a devoted experiment at JLab upgraded
to 12 GeV.

\section*{2. The model}

\subsection*{2.1. Direct non-resonant $J/\psi$ photoproduction mechanism}

  Since we are interested in the incident photon energies up to
11 GeV, which are well in the near future within the capabilities of the upgraded CEBAF facility
at JLab, we accounted for the following direct non-resonant elementary $J/\psi$ production processes
with the lowest free production threshold ($\approx$ 8.2 GeV)
:
%formula(1)
\begin{equation}
{\gamma}+p \to J/\psi+p,
\end{equation}
%formula(2)
\begin{equation}
{\gamma}+n \to J/\psi+n.
\end{equation}
Following [1], we approximate the in-medium local mass $m^*_{{J/\psi}}({\bf r})$ of the $J/\psi$ mesons,
participating in the production processes (1), (2), with their average in-medium mass
$<m^*_{{J/\psi}}>$ defined as:
%formula(3)
\begin{equation}
<m^*_{{J/\psi}}>=m_{{J/\psi}}+V_0\frac{<{\rho_N}>}{{\rho_0}}.
\end{equation}
Here, $m_{{J/\psi}}$ is the rest mass of a ${J/\psi}$ in free space, $V_0$ is the $J/\psi$ meson effective
scalar potential (or its in-medium mass shift) at normal nuclear matter density ${\rho_0}=0.16$ fm$^{-3}$,
$<{\rho_N}>$ is the average nucleon density. For target nuclei $^{12}$C
and $^{208}$Pb, considered in the present work, the ratio $<{\rho_N}>/{\rho_0}$
is approximately equal to 0.5 and 0.8, respectively. In the following study in line with [1]
for the $J/\psi$ mass shift at saturation density $V_0$ we will adopt the following
options: i) $V_0=0$, ii) $V_0=-25$ MeV, iii) $V_0=-50$ MeV, iv) $V_0=-100$ MeV, and v) $V_0=-150$ MeV
as well as will neglect the medium modification of the outgoing nucleon mass.

  Then, ignoring the distortion of the incident photon in the energy range of interest
and describing the full-sized [1] $J/\psi$ meson final-state absorption by the absorption cross section
$\sigma_{{J/\psi}N}$, we represent the total cross section for the production of ${J/\psi}$ mesons
on nuclei in the direct non-resonant processes (1) and (2) as follows [1]:
%formula(4)
\begin{equation}
\sigma_{{\gamma}A\to {J/\psi}X}^{({\rm dir})}(E_{\gamma})=I_{V}[A,\sigma_{{J/\psi}N}]
\left<\sigma_{{\gamma}N \to {J/\psi}N}(E_{\gamma})\right>_A,
\end{equation}
where
%formula(5)
\begin{equation}
I_{V}[A,\sigma]=2{\pi}A\int\limits_{0}^{R}r_{\bot}dr_{\bot}
\int\limits_{-\sqrt{R^2-r_{\bot}^2}}^{\sqrt{R^2-r_{\bot}^2}}dz
\rho(\sqrt{r_{\bot}^2+z^2})
\exp{\left[-A\sigma\int\limits_{z}^{\sqrt{R^2-r_{\bot}^2}}
\rho(\sqrt{r_{\bot}^2+x^2})dx\right]},
\end{equation}
%formula(6)
\begin{equation}
\left<\sigma_{{\gamma}N \to {J/\psi}N}(E_{\gamma})\right>_A=
\int\int
P_A({\bf p}_t,E)d{\bf p}_tdE
\sigma_{{\gamma}N \to {J/\psi}N}(\sqrt{s},<m^*_{{J/\psi}}>)
\end{equation}
and
%formula(7)
\begin{equation}
  s=(E_{\gamma}+E_t)^2-({\bf p}_{\gamma}+{\bf p}_t)^2,
\end{equation}
%formula(8)
\begin{equation}
   E_t=M_A-\sqrt{(-{\bf p}_t)^2+(M_{A}-m_{N}+E)^{2}}.
\end{equation}
Here,
$\sigma_{{\gamma}N\to {J/\psi}N}(\sqrt{s},<m^*_{{J/\psi}}>)$ is the "in-medium"
total cross section for the production of ${J/\psi}$
with reduced mass $<m^*_{{J/\psi}}>$ in reactions (1) and (2)
at the ${\gamma}N$ center-of-mass energy $\sqrt{s}$;
$\rho({\bf r})$ and $P_A({\bf p}_t,E)$ are the local nucleon density and the
spectral function of target nucleus $A$ normalized to unity
(the specific information about these quantities, used in our calculations, is given in [24, 25]);
${\bf p}_{t}$  and $E$ are the internal momentum and binding energy of the struck target nucleons
involved in the collision processes (1) and (2); $A$ is the number of nucleons in
the target nucleus, $M_{A}$  and $R$ are its mass and radius; $m_N$ is the free space nucleon mass;
${\bf p}_{\gamma}$ and $E_{\gamma}$ are the momentum and energy of the initial photon. For the
$J/\psi$--nucleon absorption cross section $\sigma_{{J/\psi}N}$ we have used the value
$\sigma_{{J/\psi}N}=3.5$ mb motivated by the results from the $J/\psi$ photoproduction experiment
at SLAC [26, 27]
\footnote{$^)$It should be pointed out that this value of the ${J/\psi}N$ inelastic cross section lies
between those of the order of 2.0--2.5 mb and 6--8 mb predicted, respectively, in Ref. [28] and Ref. [29]
using the boson exchange model and effective Lagrangians for $J/\psi$ momenta in the lab frame
$\sim$ 4--10 GeV/c. As shown in our calculations, at these momenta $J/\psi$ mesons are produced for 
photon energies studied in the present work. Therefore, in view of the fact that the $J/\psi$
meson momentum was not explicitly fixed by the analysis in Refs. [26, 27], the adopting of the cross
section $\sigma_{{J/\psi}N}=3.5$ mb in our calculations is quite reasonable.}$^)$
.

  As before in [1], we assume that the "in-medium" cross section
$\sigma_{{\gamma}N \to {J/\psi}N}({\sqrt{s}},<m^*_{J/\psi}>)$ for $J/\psi$ production in reactions (1) and (2)
is equivalent to the vacuum cross section $\sigma_{{\gamma}N \to {J/\psi}N}({\sqrt{s}},m_{J/\psi})$ in which
the free mass $m_{J/\psi}$ is replaced by the average in-medium mass $<m^*_{{J/\psi}}>$ as given by
equation (3) and the free space center-of-mass energy squared s, presented by the formula (10) below,
is replaced by the in-medium expression (7).
For the free total cross section $\sigma_{{\gamma}N \to {J/\psi}N}({\sqrt{s}},m_{J/\psi})$
in the photon energy range $E_{\gamma} \le $ 22 GeV we have used the following parametrization [1]:
%formula(9)
\begin{equation}
\sigma_{{\gamma}N \to {J/\psi}N}({\sqrt{s}},m_{J/\psi})=11.1(1-x)^2~[{\rm nb}],
\end{equation}
where
%formula(10)
\begin{equation}
  x=(s_{\rm thr}-m^2_N)/(s-m^2_N)=E^{\rm thr}_{\gamma}/E_{\gamma}, \,\,\,\,
  s_{\rm thr}=(m_{J/\psi}+m_N)^2, \,\,\,\,
  s=(E_{\gamma}+m_N)^2-{\bf p}_{\gamma}^2.
\end{equation}
Here, $E^{\rm thr}_{\gamma}=(s_{\rm thr}-m^2_N)/2m_N$ is the energy at kinematic threshold.

\subsection*{2.2. Two-step resonant $J/\psi$ photoproduction mechanism}

  At the initial photon beam energies of interest, an incident photon can produce a $P^+_c(4450)$
resonance directly in the first inelastic ${\gamma}p$ collision
\footnote{$^)$The threshold (resonant) energy $E^{\rm R}_{\gamma}$ for the photoproduction of this resonance
with pole mass $M_c=4.45$ GeV on a free target proton being at rest is $E^{\rm R}_{\gamma}=10.08$ GeV.}$^)$
:
%formula(11)
\begin{equation}
{\gamma}+p \to P^+_c(4450).
\end{equation}
Then the produced pentaquark resonance can decay into the $J/\psi$ and $p$:
%formula(12)
\begin{equation}
P^+_c(4450) \to J/\psi+p.
\end{equation}
Presently the branching ratio $Br[P^+_c(4450) \to {J/\psi}p]$ of this decay is not known.
Following Refs. [21--23, 6, 30], we will employ in our study for this ratio the three following realistic options:
$Br[P^+_c(4450) \to {J/\psi}p]=1$, 3 and 5\%
\footnote{$^)$It should be pointed out that the statistical analysis of [31] found that an upper limit for the
branching ratio of the $P^+_c(4450)$ pentaquark with the preferred [21--23] spin-parity combination $J^P=(5/2)^+$
to ${J/\psi}p$ is between 8--17\%.}$^)$
.
Most of the $P^+_c(4450)$'s decay to $J/\psi$ and $p$ essentially outside the target nuclei of interest.
This is due to the following. The $P^+_c(4450)$ resonance decay mean free path
can be estimated as $\lambda_{\rm dec}=p_c/(M_c{\Gamma})$,
where $p_c$ and $\Gamma$ are its laboratory momentum and vacuum total decay width in its rest frame.
For typical values $p_c=E^{\rm R}_{\gamma}=10.08$ GeV and $\Gamma=39$ MeV [2],
we get that $\lambda_{\rm dec} \approx 11.5$ fm. This value is
larger than the radii of $^{12}$C and $^{208}$Pb target nuclei, which are approximately 3 and 7.7 fm,
respectively.

   In our model the free spectral function of the pentaquark state $P^+_c(4450)$, produced in reaction (11),
is described by the non-relativistic Breit-Wigner distribution [21, 22, 32]:
%formula(13)
\begin{equation}
S_c(\sqrt{s},\Gamma)=\frac{1}{2\pi}\frac{\Gamma}{(\sqrt{s}-M_c)^2+{\Gamma}^2/4},
\end{equation}
where the total ${\gamma}p$ center-of-mass energy $\sqrt{s}$ is given by Eq.~(10). When obtaining the
$J/\psi$ creation cross section from the production/decay chain ${\gamma}p \to P^+_c(4450)$,
$P^+_c(4450) \to {J/\psi}p$ on $^{12}$C and $^{208}$Pb target nuclei in the "free" $P^+_c(4450)$
spectral function scenario (see Fig.~2 below), this energy is calculated according to Eq.~(7).
Following [33], we assume that the in-medium $P^+_c(4450)$ resonance spectral function
$S_c(\sqrt{s},\Gamma_{\rm med})$ is also described by the Breit-Wigner formula (13) with a total in-medium
width $\Gamma_{\rm med}$ in its rest frame, obtained as a sum of the vacuum decay width, $\Gamma$, and an
in-medium contribution due to $P^+_c(4450)N$ inelastic collisions -- averaged over local nucleon density
$\rho_N({\bf r})$ collisional width $<\Gamma_{\rm coll}>$
\footnote{$^)$In using (13) to describe the in-medium $P^+_c(4450)$ resonance we will neglect the possible
shift of its pole mass in it.}$^)$
:
%formula(14)
\begin{equation}
\Gamma_{\rm med}=\Gamma+<\Gamma_{\rm coll}>,
\end{equation}
where, according to [34] and in view of Eq.~(3), the average collisional width $<\Gamma_{\rm coll}>$ reads:
%formula(15)
\begin{equation}
<\Gamma_{\rm coll}>={\gamma_c}{v_c}{\sigma_{P_cN}}<\rho_N>.
\end{equation}
Here, $\sigma_{P_cN}$ is the $P^+_c(4450)$--nucleon inelastic cross section and the Lorentz $\gamma$-factor
$\gamma_c$ and the velocity $v_c$ of the resonance in the nuclear rest frame are given by:
%formula(16)
\begin{equation}
\gamma_c=\frac{(E_{\gamma}+E_t)}{\sqrt{s}},\,\,\,\,\,v_c=\frac{|{\bf p}_{\gamma}+{\bf p}_t|}{(E_{\gamma}+E_t)}.
\end{equation}
Taking into account the $P^+_c(4450)$ resonance minimum quark content ($P^+_c(4450)=c{\bar c}uud$)
as well as the fact that in the baryocharmonium model [21, 22] it is interpreted as a composite
of $J/\psi$ and excited nucleon state with the quantum numbers of resonance $N(1520)$, the cross section
$\sigma_{P_cN}$ can be evaluated as:
%formula(17)
\begin{equation}
\sigma_{P_cN} \approx \sigma_{{J/\psi}N}+\sigma_{pN}^{\rm in},
\end{equation}
where $\sigma_{pN}^{\rm in}$ is the inelastic cross section of the free $pN$ interaction.
Using $\sigma_{{J/\psi}N}=3.5$ mb [26, 27] and $\sigma_{pN}^{\rm in}=30$ mb [35], we get that
$\sigma_{P_cN}=33.5$ mb. Thus, for example, with this value as well as with $<\rho_N>=0.5\rho_0$ ($^{12}$C),
$<\rho_N>=0.8\rho_0$ ($^{208}$Pb), ${\rho_0}=0.16$ fm$^{-3}$, the expression (15) gives numerically
that $<\Gamma_{\rm coll}> \approx 120$ MeV for $^{12}$C and
$<\Gamma_{\rm coll}> \approx 190$ MeV for $^{208}$Pb in the case when the $P^+_c(4450)$ resonance
is created in the reaction channel (11) on a free target proton at rest by incident photon with energy
$E_{\gamma}=E_{\gamma}^{\rm R}=10.08$ GeV.
%%%%%%%%%%%%%%%%%%%%%%%%%%%%%%%%%%%%%%%%%%%%%%%%%%%%%%%%%%%
\begin{figure}[htb]
\begin{center}
\includegraphics[width=16.0cm]{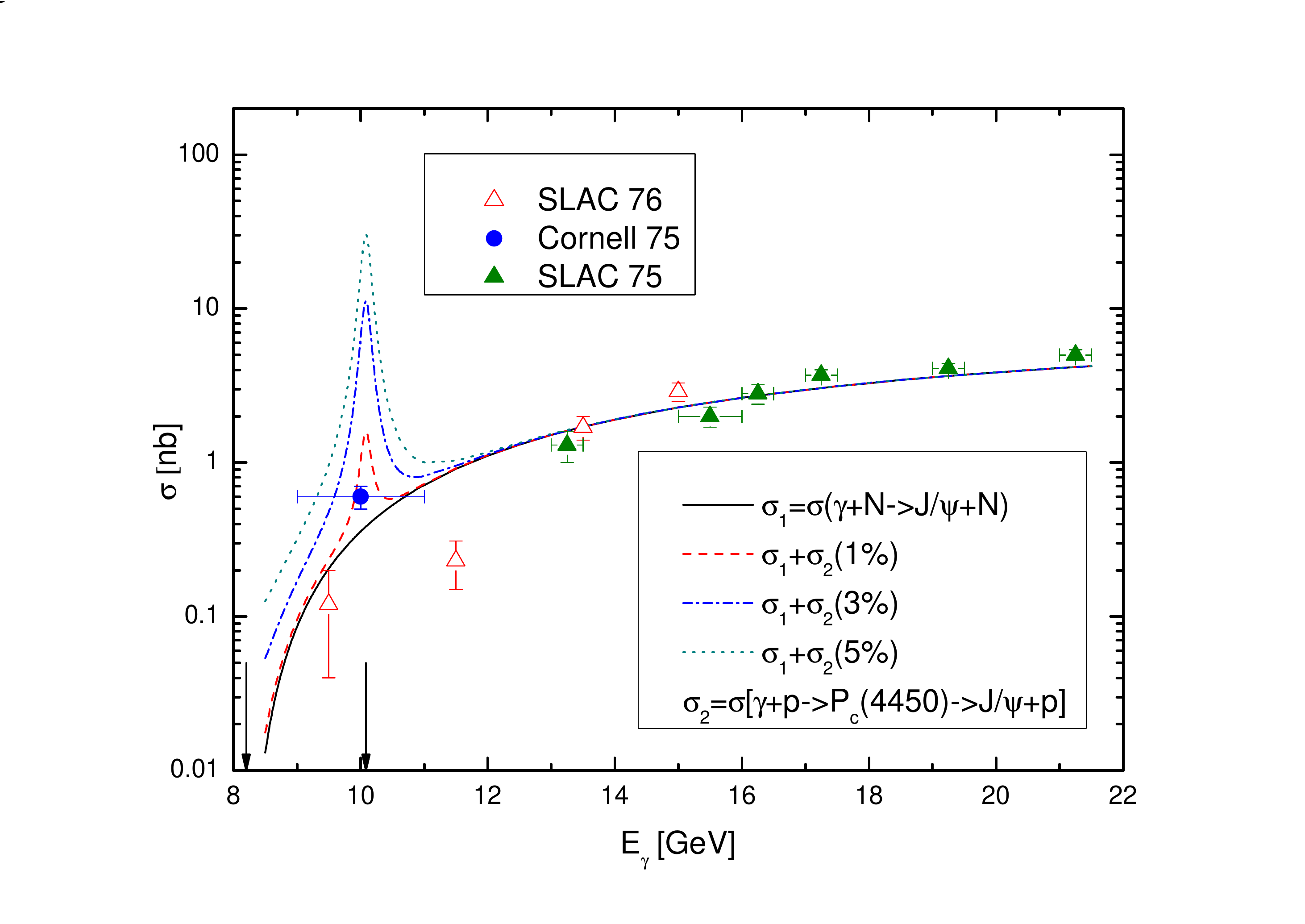}
\vspace*{-2mm} \caption{(color online) The non-resonant total cross section for the reaction
${\gamma}N \to {J/\psi}N$ (solid curve) and the incoherent sum of it and the total cross section for the
resonant $J/\psi$ production in the process ${\gamma}p \to P^+_c(4450) \to {J/\psi}p$, calculated assuming
that the resonance $P^+_c(4450)$ with the spin-parity quantum numbers $J^P=(5/2)^+$ decays to ${J/\psi}p$
with the lower allowed relative orbital angular momentum $L=1$ with branching fractions
$Br[P^+_c(4450) \to {J/\psi}p]=1$, 3 and 5\% (respectively, dashed, dotted-dashed and dotted curves),
as functions of photon energy. The left and right arrows indicate, correspondingly, the threshold energy
$E^{\rm thr}_{\gamma}=8.2$ GeV for the reaction ${\gamma}N \to {J/\psi}N$ proceeding on a free target
nucleon being at rest and the resonant energy $E^{\rm R}_{\gamma}=10.08$ GeV.}
\label{void}
\end{center}
\end{figure}
%%%%%%%%%%%%%%%%%%%%%%%%%%%%%%%%%%%%%%%%%%%%%%%%%%%%%%%%%%%%%%%%%%%%%%%%%%%%%%%%%%%%%%%%%%%%%

  The free Breit-Wigner total cross section for production of a $P^+_c(4450)$ resonance with spin $J=5/2$
in reaction (11) can be described on the basis of the spectral function (13), provided that the branching
ratio $Br[P^+_c(4450) \to {\gamma}p]$ is known, as follows [21, 22, 32, 36]:
%formula(18)
\begin{equation}
\sigma_{{\gamma}p \to P^+_c(4450)}(\sqrt{s},\Gamma)=\frac{3\pi^2}{k^2}Br[P^+_c(4450) \to {\gamma}p]
S_c(\sqrt{s},\Gamma)\Gamma,
\end{equation}
where the factor 3 appears from the ratio of spin factors and $k$ denotes the center-of-mass momentum
in the incoming ${\gamma}p$ channel. Employing the proper kinematics, we obtain that
%formula(19)
\begin{equation}
k=\frac{1}{2\sqrt{s}}\lambda(s,0,m_N^2),
\end{equation}
where
%FORMULA (20)
\begin{equation}
\lambda(x,y,z)=\sqrt{{\left[x-({\sqrt{y}}+{\sqrt{z}})^2\right]}{\left[x-
({\sqrt{y}}-{\sqrt{z}})^2\right]}}.
\end{equation}
When the reaction (11) occurs on an off-shell target proton then instead of the nucleon mass squared $m_N^2$
in Eq.~(19) we should use the quantity $E_t^2-p_t^2$. It should be calculated in line with Eq.~(8).
Following [31], we assume that the $P^+_c(4450)$ $(5/2)^+$ decay to ${J/\psi}p$ is dominated by the lowest
partial wave with relative orbital angular momentum $L=1$. In this case the branching
ratio $Br[P^+_c(4450) \to {\gamma}p]$ can be expressed via the branching fraction
$Br[P^+_c(4450) \to {J/\psi}p]$ in the following manner [21, 22, 31, 32]:
%formula(21)
\begin{equation}
Br[P^+_c(4450) \to {\gamma}p]=7.5\cdot10^{-3}Br[P^+_c(4450) \to {J/\psi}p].
\end{equation}
Within the representation of Eq.~(18), the free resonant total cross section
$\sigma_{{\gamma}p \to P^+_c(4450)\to {J/\psi}p}(\sqrt{s},\Gamma)$ for $J/\psi$ production in the
two-step process (11), (12) can be written in the following form:
%FORMULA (22)
\begin{equation}
\sigma_{{\gamma}p \to P^+_c(4450)\to {J/\psi}p}(\sqrt{s},\Gamma)=
\sigma_{{\gamma}p \to P^+_c(4450)}(\sqrt{s},\Gamma)\theta[\sqrt{s}-(m_{J/\psi}+m_N)]
Br[P^+_c(4450) \to {J/\psi}p].
\end{equation}
Here, $\theta(x)$ is the standard step function.
In view of Eqs.~(18) and (21) this cross section is proportional to $Br^2[P^+_c(4450) \to {J/\psi}p]$.

  The size of the non-resonant (9) and resonant (22) cross sections in comparison with the scarce existing
low-energy data from Cornell 75 [37] (full dot), SLAC 75 [38] (full triangles) and SLAC 76 [39]
(open triangles), collected together in [40], is illustrated by  Fig.~1. It can be seen that at photon
energies around the peak energy $E^{\rm R}_{\gamma}=10.08$ GeV the non-resonant $J/\psi$ production cross
section is small compared to the resonant contribution. The combined (non-resonant plus resonant) total
cross section (dashed, dotted-dashed and dotted curves) has here a non-trivial behavior due to the
excitation of the $P^+_c(4450)$ resonance and its peak values reach tens of nanobarns, if
$Br[P^+_c(4450) \to {J/\psi}p]$ $\sim$ 3--5\%. This cross section, calculated for
$Br[P^+_c(4450) \to {J/\psi}p]=1$\% (dashed curve), has a peak value of about 2 nb, which is consistent
with that obtained before in Ref. [23] for the coherent sum of the contributions to the total cross section
of ${\gamma}p \to {J/\psi}p$ reaction from the $t$-channel diffractive Pomeron exchange and the $s$-channel
pentaquark $P^+_c(4450)$ production with the quantum numbers $J^P=(5/2)^+$, but assuming that the ${J/\psi}p$
decay channel accounts for 5\% of its total width. The latter two peak values for the combined total cross
section are compatible with the smeared by poor photon energy
resolution Cornell 75 $J/\psi$ photoproduction data point [37]. In this context, a detailed scan of the
$J/\psi$ total photoproduction cross section on a proton target in the near-threshold energy region around
$E^{\rm R}_{\gamma}=10.08$ GeV in future experiments at JLab should help us to obtain a definite result
for or against the existence of the genuine $P^+_c(4450)$ pentaquark state and to clarify its nature and decay
probabilities.
%%%%%%%%%%%%%%%%%%%%%%%%%%%%%%%%%%%%%%%%%%%%%%%%%%%%%%%%%%%
\begin{figure}[!h]
\begin{center}
\includegraphics[width=16.0cm]{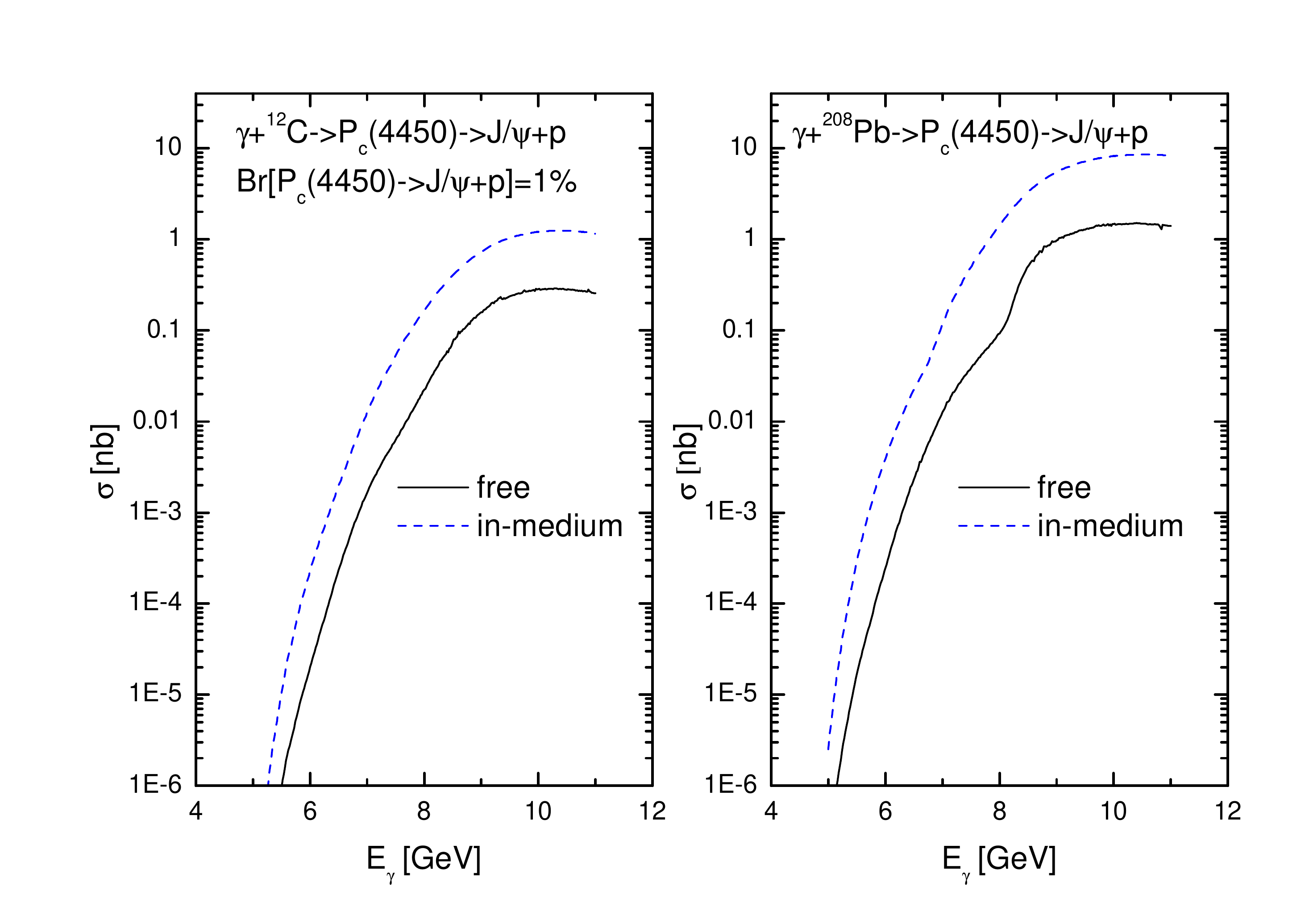}
\vspace*{-2mm} \caption{(color online) Excitation functions for resonant production of $J/\psi$
mesons off $^{12}$C and $^{208}$Pb from the process ${\gamma}p \to P^+_c(4450) \to {J/\psi}p$ going on an
off-shell target proton. The curves are calculations for $Br[P^+_c(4450) \to {J/\psi}p]=1$\% adopting free
(solid curves) and in-medium (dashed curves) $P^+_c(4450)$ spectral functions as described in the text.}
\label{void}
\end{center}
\end{figure}
%%%%%%%%%%%%%%%%%%%%%%%%%%%%%%%%%%%%%%%%%%%%%%%%%%%%%%%%%%%%%%%%%%%%%%%%%%%%%%%%%%%%%%%%%%%%%
%%%%%%%%%%%%%%%%%%%%%%%%%%%%%%%%%%%%%%%%%%%%%%%%%%%%%%%%%%%
\begin{figure}[!h]
\begin{center}
\includegraphics[width=16.0cm]{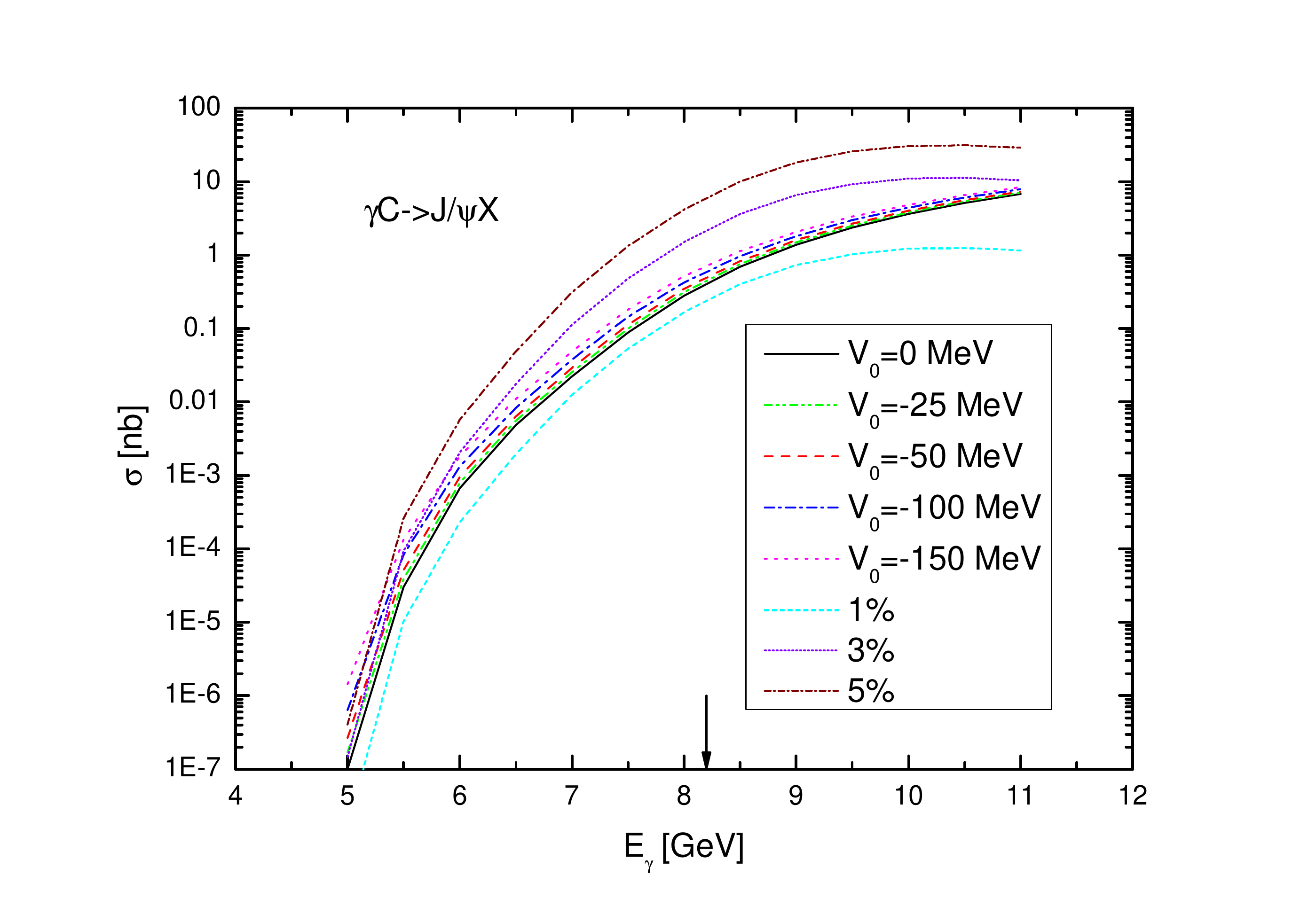}
\vspace*{-2mm} \caption{(color online) Excitation functions for the non-resonant and resonant production
of $J/\psi$ mesons off $^{12}$C from direct ${\gamma}N \to {J/\psi}N$ and resonant
${\gamma}p \to P^+_c(4450) \to {J/\psi}p$ reactions going on an off-shell target nucleons. The curves,
corresponding to the non-resonant production of $J/\psi$ mesons, are calculations with an in-medium
$J/\psi$ mass shift depicted in the inset. The curves, belonging to their resonant production, are
calculations for $Br[P^+_c(4450) \to {J/\psi}p]=1$, 3 and 5\% adopting
in-medium $P^+_c(4450)$ spectral function as described in the text. The arrow indicates the threshold
energy for direct $J/\psi$ photoproduction on a free target nucleon being at rest.}
\label{void}
\end{center}
\end{figure}
%%%%%%%%%%%%%%%%%%%%%%%%%%%%%%%%%%%%%%%%%%%%%%%%%%%%%%%%%%%
%%%%%%%%%%%%%%%%%%%%%%%%%%%%%%%%%%%%%%%%%%%%%%%%%%%%%%%%%%%
\begin{figure}[!h]
\begin{center}
\includegraphics[width=16.0cm]{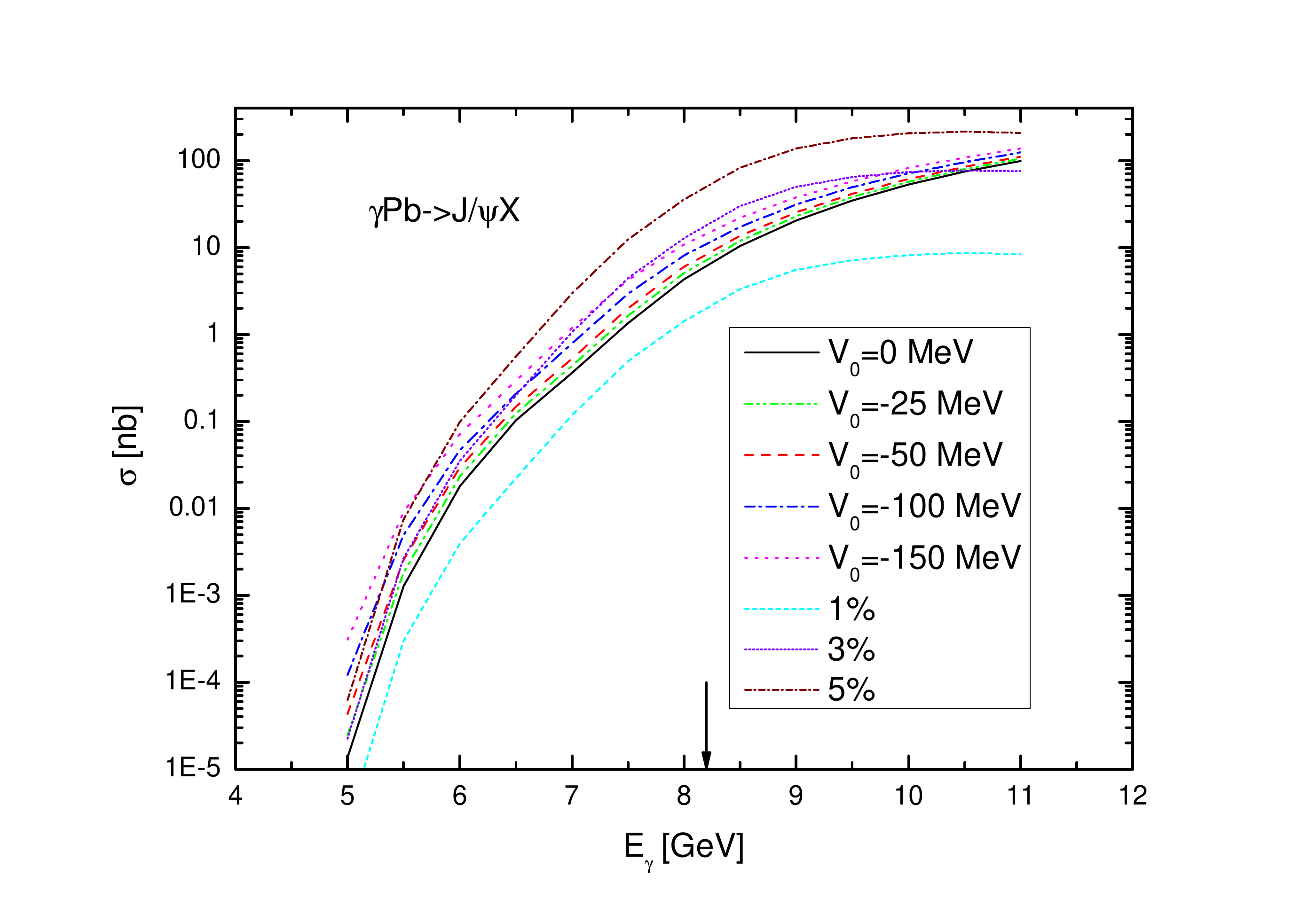}
\vspace*{-2mm} \caption{(color online) The same as in figure 3, but for the $^{208}$Pb target nucleus.}
\label{void}
\end{center}
\end{figure}
%%%%%%%%%%%%%%%%%%%%%%%%%%%%%%%%%%%%%%%%%%%%%%%%%%%%%%%%%%%
%%%%%%%%%%%%%%%%%%%%%%%%%%%%%%%%%%%%%%%%%%%%%%%%%%%%%%%%%%%
\begin{figure}[!h]
\begin{center}
\includegraphics[width=16.0cm]{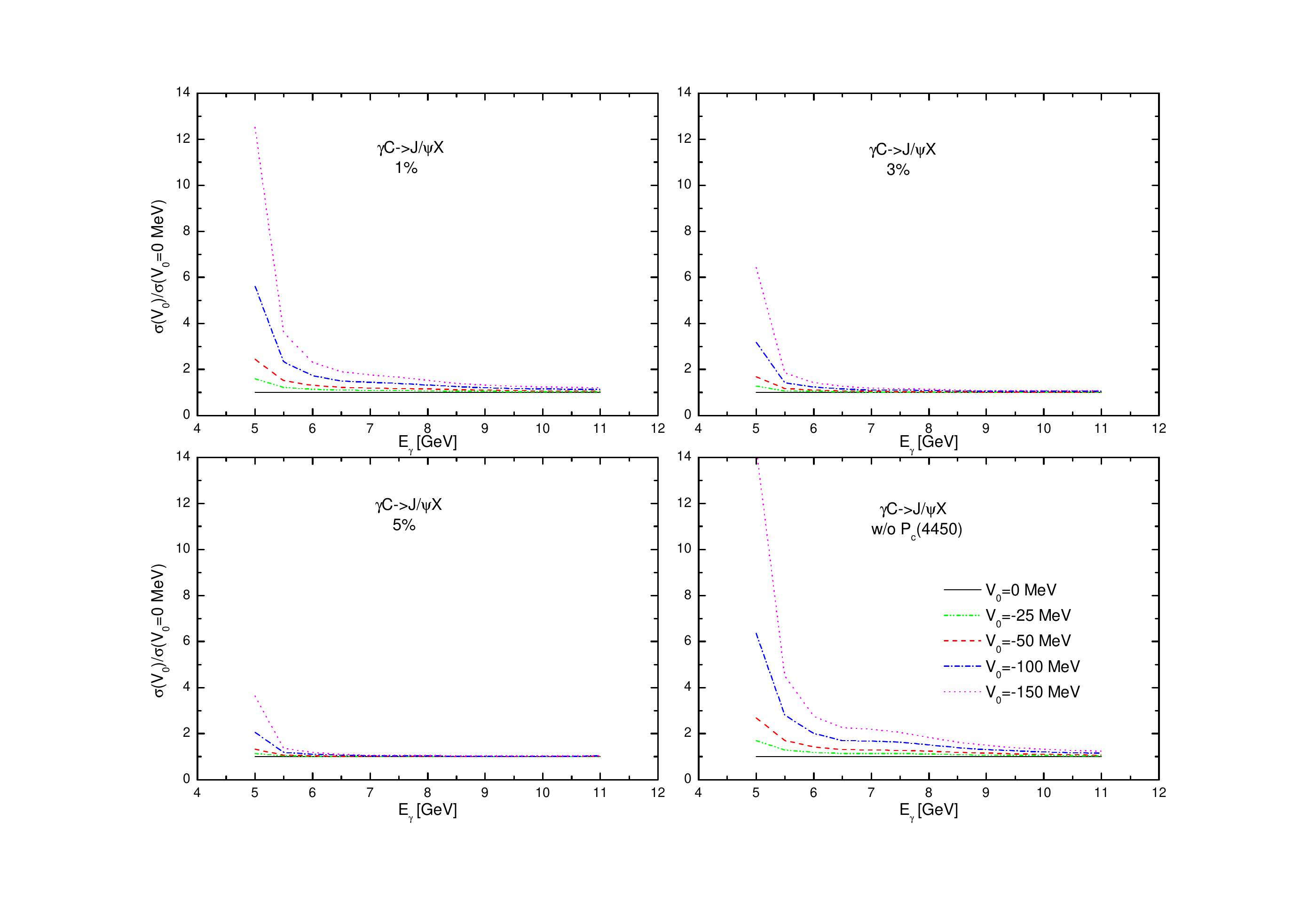}
\vspace*{-2mm} \caption{(color online) Ratio between the sum of the cross sections for the non-resonant
and resonant photoproduction of $J/\psi$ off $^{12}$C, calculated, respectively, with the employed $J/\psi$
mass shift, with values shown in the inset,
and $P^+_c(4450)$ branching fractions to ${J/\psi}p$ of 1, 3 and 5\% and presented in figure 3,
and the sum of the same cross sections, obtained, correspondingly, without this shift and with the same
$P^+_c(4450)$ branching fractions of 1\% (upper left panel), 3\% (upper right panel) and 5\% (lower left panel),
as a function of photon energy. The same as above, but only determined without the contribution to the $J/\psi$
photoproduction from the resonant process ${\gamma}p \to P^+_c(4450) \to {J/\psi}p$ (lower right panel).}
\label{void}
\end{center}
\end{figure}
%%%%%%%%%%%%%%%%%%%%%%%%%%%%%%%%%%%%%%%%%%%%%%%%%%%%%%%%%%%
%%%%%%%%%%%%%%%%%%%%%%%%%%%%%%%%%%%%%%%%%%%%%%%%%%%%%%%%%%%
\begin{figure}[!h]
\begin{center}
\includegraphics[width=16.0cm]{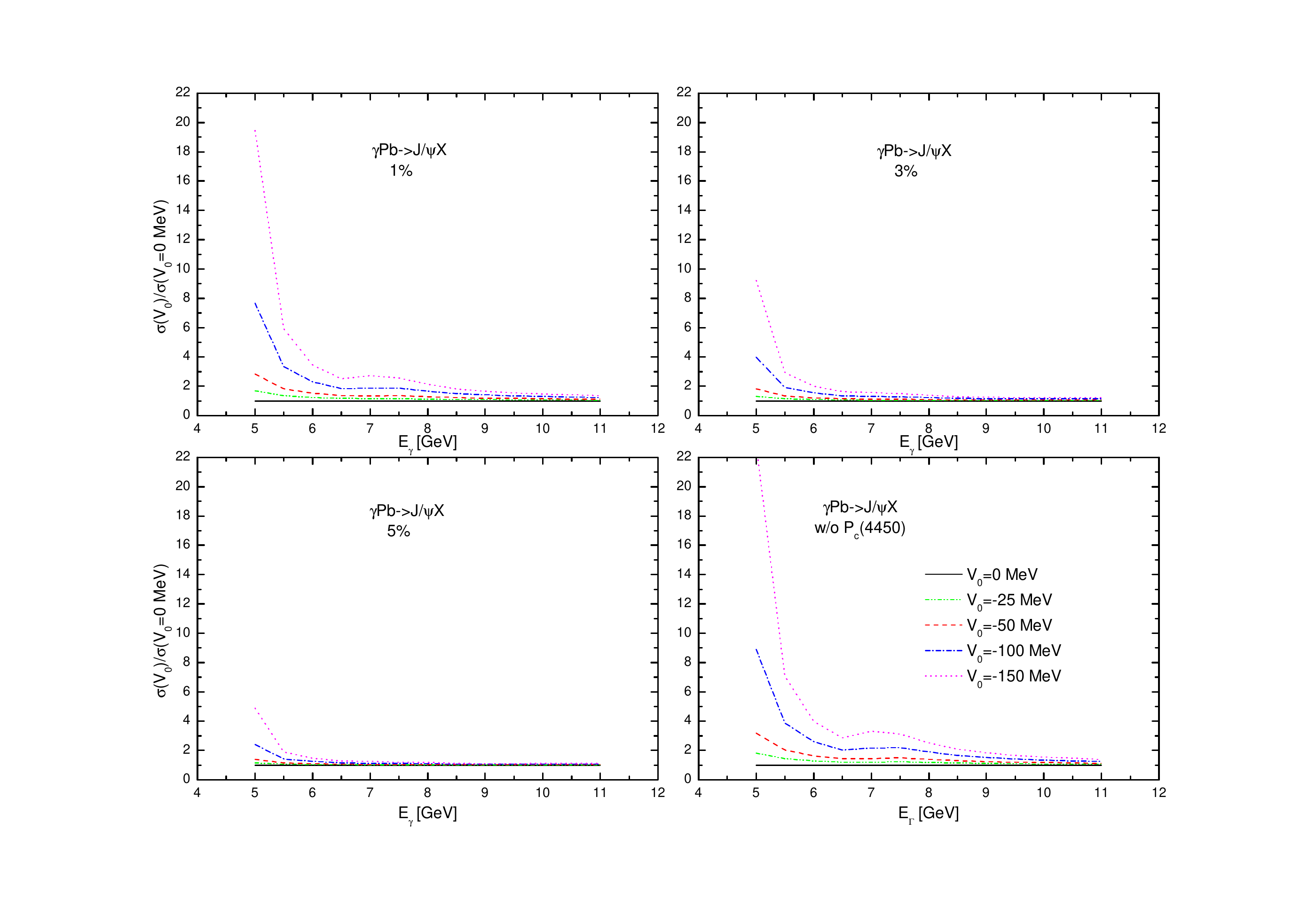}
\vspace*{-2mm} \caption{(color online) Ratio between the sum of the cross sections for the non-resonant
and resonant photoproduction of $J/\psi$ off $^{208}$Pb, calculated, respectively, with the employed $J/\psi$
mass shift, with values shown in the inset,
and $P^+_c(4450)$ branching fractions to ${J/\psi}p$ of 1, 3 and 5\% and presented in figure 4,
and the sum of the same cross sections, obtained, correspondingly, without this shift and with the same
$P^+_c(4450)$ branching fractions of 1\% (upper left panel), 3\% (upper right panel) and 5\% (lower left panel),
as a function of photon energy. The same as above, but only determined without the contribution to the $J/\psi$
photoproduction from the resonant process ${\gamma}p \to P^+_c(4450) \to {J/\psi}p$ (lower right panel).}
\label{void}
\end{center}
\end{figure}
%%%%%%%%%%%%%%%%%%%%%%%%%%%%%%%%%%%%%%%%%%%%%%%%%%%%%%%%%%%

   Taking into account the fact that the most of the produced $P^+_c(4450)$ resonances decay to $J/\psi$
and $p$ outside the target nuclei considered as well as using the results presented above by Eqs.~(4)--(8)
and those given in Ref.~[41], we get the following expression for the $J/\psi$ total cross section for
${\gamma}A$ interactions from the production/decay chain (11), (12):
%formula(23)
\begin{equation}
\sigma_{{\gamma}A\to P^+_c(4450)\to{J/\psi}p}^{({\rm sec})}(E_{\gamma})=\left(\frac{Z}{A}\right)
I_{V}[A,\sigma^{\rm eff}_{P_cN}]\left<\sigma_{{\gamma}p \to P^+_c(4450)}(E_{\gamma})\right>_A
Br[P^+_c(4450) \to {J/\psi}p],
\end{equation}
where
%formula(24)
\begin{equation}
\left<\sigma_{{\gamma}p \to P^+_c(4450)}(E_{\gamma})\right>_A=
\int\int
P_A({\bf p}_t,E)d{\bf p}_tdE
\sigma_{{\gamma}p \to P^+_c(4450)}(\sqrt{s},\Gamma_{\rm med})\theta[\sqrt{s}-(m_{J/\psi}+m_N)].
\end{equation}
Here, $\sigma_{{\gamma}p \to P^+_c(4450)}(\sqrt{s},\Gamma_{\rm med})$ is the "in-medium" cross section
for the $P^+_c(4450)$ resonance production in a ${\gamma}p$ collision (11), $Z$ is the number of protons
in the target nucleus and $\sigma^{\rm eff}_{P_cN}$ is the $P^+_c(4450)$--nucleon effective absorption
cross section, which will be defined below. The quantity $I_{V}[A,\sigma^{\rm eff}_{P_cN}]$ in Eq.~(23)
is determined by Eq.~(5) in which one has to make the substitution $\sigma \to \sigma^{\rm eff}_{P_cN}$.
As above in Eq.~(6), we assume that the "in-medium" cross section
$\sigma_{{\gamma}p \to P^+_c(4450)}(\sqrt{s},\Gamma_{\rm med})$ is equivalent to the free cross section
$\sigma_{{\gamma}p \to P^+_c(4450)}(\sqrt{s},\Gamma)$ of Eq.~(18) in which the vacuum decay width $\Gamma$
is replaced by the in-medium width $\Gamma_{\rm med}$ as given by Eqs.~(14)--(16) and the free space
center-of-mass energy squared $s$, presented by the expression (10), is replaced by the in-medium formula
(7). Moreover, the branching ratio $Br[P^+_c(4450) \to {\gamma}p]$, entering into the Eq.~(18), is assumed
to be independent on the nuclear medium.

 Let us focus now on the effective absorption cross section $\sigma^{\rm eff}_{P_cN}$, which governs the
survival probability
%formula(25)
\begin{equation}
\exp{\left[-A\sigma^{\rm eff}_{P_cN}\int\limits_{z}^{\sqrt{R^2-r_{\bot}^2}}
\rho(\sqrt{r_{\bot}^2+x^2})dx\right]}
\end{equation}
of the $P^+_c(4450)$ resonance in Eqs.~(5) and (23) until it is escaped to the vacuum.
The attenuation of the $P^+_c(4450)$ flux in the nucleus is caused both by the inelastic
$P^+_c(4450)$--nucleon collisions and by $P^+_c(4450)$ decays here. In line with (14), these
decays are determined by the free decay width $\Gamma$. In the low-density approximation (15),
this width is related to an additional to the inelastic cross section $\sigma_{P_cN}$
effective $P^+_c(4450)$ absorption cross section $\sigma_{\rm dec}$. Using $\Gamma=39$ MeV,
we obtain that $\sigma_{\rm dec}\approx 10$ mb for target nuclei considered in the case of
the free $P^+_c(4450)$ state production on a target proton at rest by incident photon with resonant
energy $E_{\gamma}^{\rm R}=10.08$ GeV. In line with above mentioned, the cross section
$\sigma^{\rm eff}_{P_cN}$ can be represented as follows:
%formula(26)
\begin{equation}
\sigma^{\rm eff}_{P_cN}=\sigma_{P_cN}+\sigma_{\rm dec}.
\end{equation}
Therefore, with values $\sigma_{\rm dec}\approx 10$ mb and $\sigma_{P_cN} \approx 33.5$ mb,
the expression (26) gives that
$\sigma^{\rm eff}_{P_cN}\approx 43.5$ mb. We will employ this value in our calculations of the cross
section of $P^+_c(4450)$ production in ${\gamma}A$ collisions.

\section*{3. Results}

 Figure 2 shows the energy dependence of the total resonant $J/\psi$ production
cross section for ${\gamma}$C and $\gamma$Pb collisions. The $J/\psi$ production is calculated
on the basis of Eqs.~(23), (24) in the scenarios with free and in-medium $P^+_c(4450)$ spectral
functions for branching ratio $Br[P^+_c(4450) \to {J/\psi}p]=1$\%. One can see that the $P^+_c(4450)$
resonance formation is rather smeared out by Fermi motion of intranuclear protons at photon energies
around 10.08 GeV (cf. Fig.~1) and it is a sizeably enhanced at all incident energies considered for
the in-medium case.

    Excitation functions for direct production of $J/\psi$ mesons as well as for their resonant creation
via $P^+_c(4450)$ resonance formation and decay in case of ${\gamma}$C and $\gamma$Pb interactions are
displayed in Figs.~(3) and (4), respectively. The former ones are calculated on the basis of Eq.~(4)
for five employed options for the $J/\psi$ in-medium mass shift (cf. [1]), whereas the latter functions
are obtained using Eq.~(23) in the in-medium $P^+_c(4450)$ spectral function scenario and assuming that
branching ratio $Br[P^+_c(4450) \to {J/\psi}p]=1$, 3 and 5\%. It is seen that in the far subthreshold
energy region ($E_{\gamma}$ $\sim$ 5--7 GeV), where the influence of the charmonium mass shift on its
non-resonant yield is essential, this yield (for considered options for the mass shift) and that from the
production and decay of the intermediate $P^+_c(4450)$ resonance are comparable for
$Br[P^+_c(4450) \to {J/\psi}p]=3$ and 5\%. At above threshold photon energies of 8.2--11.0 GeV
the resonant $J/\psi$ production cross section on $^{12}$C, calculated for these values of the
branching ratio $Br[P^+_c(4450) \to {J/\psi}p]$, and that on $^{208}$Pb, obtained for
$Br[P^+_c(4450) \to {J/\psi}p]=5$\%, are larger than the corresponding non-resonant ones.
If $Br[P^+_c(4450) \to {J/\psi}p]=1$\%, then the situation is vice versa. Thus, the presence of the
$P^+_c(4450)$ resonance in $J/\psi$ photoproduction, on the one hand, produces above threshold additional
enhancements in the behavior of the total $J/\psi$ creation cross section on nuclei, which could be studied
at JLab in the near future to provide further evidence for its existence. On the other hand, such presence masks
the modification of the mass of the directly photoproduced $J/\psi$ mesons in nuclear medium and, therefore,
makes the determination of this modification from the excitation function measurements at far subthreshold beam energies ($E_{\gamma}$ $\sim$ 5--7 GeV) difficult, if $Br[P^+_c(4450) \to {J/\psi}p]$ $\sim$ 3--5\% and more.

  To see more clearly the sensitivity of the overall $J/\psi$ meson yield to its in-medium mass shift,
in Figs.~5 and 6 we show on a linear scale the ratios between the sum of the non-resonant $J/\psi$
production cross section, calculated with different values for the mass shift at saturation density, as well as
resonant one, obtained for branching ratios to ${J/\psi}p$ of 1, 3, 5\%, and the same sum, determined without
this shift and with the same $P^+_c(4450)$ branching fractions to ${J/\psi}p$. Also here the same ratios as above,
but only determined without the contribution to $J/\psi$ photoproduction from the resonant process
${\gamma}p \to P^+_c(4450) \to {J/\psi}p$, are shown. It can be seen indeed that the apparent possibility of
studying the mass shift of the directly photoproduced in the nuclear matter $J/\psi$ mesons in the case of the
presence of $P^+_c(4450)$ resonance in $J/\psi$ photoproduction from the excitation function measurements exists
only in the range mass shifts of -50 MeV and more at far subthreshold photon energies
$E_{\gamma}$ $\sim$ 5.5--7.0 GeV, where the respective $J/\psi$ production cross sections on nuclei are yet
not so small (see Figs.~3 and 4), if $Br[P^+_c(4450) \to {J/\psi}p]=1$\% and less.

   Accounting for the above considerations, we come to the conclusion that the $J/\psi$ excitation function
measurements at initial photon energies below the production threshold on the free target nucleon will allow
to shed light on the possible charmonium mass shifts in the range of -50 MeV and more only if branching ratio
$Br[P^+_c(4450) \to {J/\psi}p]$ $\sim$ 1\% and less.

\section*{4. Conclusions}

 In this paper we have calculated the absolute and relative
excitation functions for $J/\psi$ production off $^{12}$C and $^{208}$Pb target nuclei at
near-threshold incident photon energies of 5--11 GeV by considering incoherent direct (${\gamma}N \to {J/\psi}N$)
and two-step (${\gamma}p \to P^+_c(4450)$, $P^+_c(4450) \to {J/\psi}p$)
photon--nucleon charmonium production processes in the framework of a nuclear spectral function approach.
It was shown that the overall $J/\psi$ production off nuclei is well sensitive to the possible
$J/\psi$ in-medium mass shifts in the range of -50 MeV and more at subthreshold beam energies only if
branching ratio $Br[P^+_c(4450) \to {J/\psi}p]$ $\sim$ 1\% and less. It was also found that the presence of the
$P^+_c(4450)$ resonance in $J/\psi$ photoproduction produces above threshold additional enhancements in the
behavior of the total $J/\psi$ creation cross section on nuclei, which could be also studied in the future
JLab experiments at CEBAF to provide further evidence for its existence.
\\

%%%%%%%%%%%%%%%%%%%%%%%%%%%%%%%%%%%%%%%%%%%%%%%%%%%%%%%%%%%%%%%%
\end{document}